\documentclass[pdflatex,sn-vancouver,Numbered]{sn-jnl}

\usepackage{graphicx}%
\usepackage{caption}
\usepackage{multirow}%
\usepackage{amsmath,amssymb,amsfonts}%
\usepackage{amsthm}%
\usepackage{mathrsfs}%
\usepackage{mathpazo}
\usepackage[title]{appendix}%
\usepackage{xcolor}%
\usepackage{textcomp}%
\usepackage{manyfoot}%
\usepackage{booktabs}%
\usepackage{algorithm}%
\usepackage{algorithmicx}%
\usepackage{algpseudocode}%
\usepackage{listings}%
\usepackage{booktabs}
\usepackage{graphicx}
\usepackage{adjustbox}
\usepackage{graphics}
\usepackage{xcolor}
\usepackage{soul}
\usepackage{lmodern}

\begin{document}

\title[Malaria forecasting]{Forecasting Malaria in Indian States: A Time Series Approach with R Shiny Integration
}


\author[1]{\fnm{Sujit K.} \sur{Ghosh}}\email{sujit\_ghosh@ncsu.edu}

\author*[2, 3]{\fnm{Usha} \sur{Ananthakumar}}\email{usha@som.iitb.ac.in}

\author[3]{\fnm{Praveen D.} \sur{Chougale}}\email{22d1629@iitb.ac.in}

\author[3]{\fnm{Adithya B.} \sur{Somaraj}}\email{absomaraj@gmail.com}

\affil[1]{\orgdiv{Department of Statistics}, \orgname{North Carolina State University}, \orgaddress{\city{Raleigh}, \postcode{27606}, \state{North Carolina}, \country{United States of America}}}

\affil[2]{\orgdiv{Shailesh J. Mehta School of Management}, \orgname{Indian Institute of Technology Bombay}, \orgaddress{\city{Mumbai}, \postcode{400076}, \state{Maharashtra}, \country{India}}}

\affil[3]{\orgdiv{Koita Centre for Digital Health}, \orgname{Indian Institute of Technology Bombay}, \orgaddress{\city{Mumbai}, \postcode{400076}, \state{Maharashtra}, \country{India}}}



\abstract{Malaria remains a significant public health challenge in many regions, necessitating robust predictive models to aid in its management and prevention. This study focuses on developing and evaluating time series models for forecasting malaria cases across eight Indian states: Jharkhand, Chhattisgarh, Maharashtra, Meghalaya, Mizoram, Odisha, Tripura, and Uttar Pradesh. We employed various modeling approaches, including polynomial regression with seasonal components, log-transformed polynomial regression, lagged difference models, and ARIMA models, to capture the temporal dynamics of malaria incidence. Comprehensive model fitting, residual analysis, and performance evaluation using metrics such as Root Mean Squared Error (RMSE) and Mean Absolute Percentage Error (MAPE) indicated that the log-transformed polynomial regression model consistently outperformed other models in terms of accuracy and robustness across all states. Rolling forecast validation further confirmed the superior predictive capability of the log-transformed model over time. Additionally, an interactive R Shiny tool was developed to facilitate the use of these predictive models by researchers and public health officials. This tool allows users to input data, select modeling approaches, and visualize predictions and performance metrics, providing a practical tool for real-time malaria forecasting and decision-making support. Our findings highlight the critical role of appropriate modeling techniques in malaria prediction and offer valuable resources for enhancing malaria surveillance and response efforts.}




\maketitle

\section{Introduction}\label{sec1}
Malaria is a major public health concern, particularly in tropical and subtropical regions, where it causes significant morbidity and mortality. According to the World Health Organization (WHO), there were an estimated 229 million cases of malaria globally in 2019, with the disease responsible for approximately 409,000 deaths, predominantly in sub-Saharan Africa and South Asia. In India, malaria remains endemic in several states, posing a continuous challenge to healthcare systems.

Many of the countries in the WHO South East Asia Region are close to achieving or have achieved complete elimination of malaria. Maldives and Sri Lanka have eradicated malaria, while Thailand, Timor-Leste, Bhutan, and Nepal aim to do so by 2025 \citep{WHO2022}. WHO has set the target for a 90\% reduction of the global malaria burden and no new indigenous cases by 2030 \citep{WHO2015}, but many countries are still far from reaching this target due to a combination of various socioeconomic, ecological, and health system-related factors. 

{India accounts for an estimated 1.7 million malaria cases in 2022} \citep{WHO2023} {with transmission intensity and species prevalence varying significantly across states and districts.} The country faces many challenges in the planning of malaria control and prevention strategies and timelines due to the malaria epidemiology in India being complicated across diverse demography, topography, and socio-cultural landscapes \citep{IndiaFramework2016}. The country has put into effect the National Strategic Plan for Malaria Elimination (2017–2022) \citep{IndiaPlan2017} and the National Framework for Malaria Elimination (NFME) (2016–2030) \citep{IndiaFramework2016}, which outline strategies such as early diagnosis and prompt treatment, vector control, community involvement, and inter-sectoral collaboration. These efforts are directed toward achieving the national goal of eliminating malaria by 2030, with particular emphasis on 27 high-priority districts where malaria transmission is moderate to high.

{Despite improvements in India’s malaria surveillance and national reporting, granular district-level data are not uniformly available or updated in real time, and inconsistencies in reporting quality remain a challenge, particularly in remote or under-resourced areas} \citep{Dhamnetiya2015, Ghosh2019}. {Moreover, publicly accessible datasets often exclude detailed epidemiological variables or lag behind actual transmission timelines, limiting their use for proactive public health interventions. Given these limitations, state-level forecasting serves as a practical intermediate resolution. It offers a balance between spatial specificity and data availability while supporting public health decision-making at the level where most operational vector control and resource allocation decisions are made.} Effective malaria control and prevention strategies rely heavily on the ability to accurately predict outbreaks and trends in malaria incidence. Traditional surveillance methods often fall short in providing timely and precise forecasts, thereby necessitating the development of advanced predictive models.

Different methods \citep{Yu2005, Zinszer2015} have been proposed for malaria forecasting to address various challenges. These can be categorized into three categories: mathematical methods \citep{MacDonald1957, Zinszer2012}, machine learning methods \citep{Zinszer2012, Anderson1995}, and statistical methods \citep{Zinszer2012, Hyndman2018}. Among these, statistical methods are particularly prominent in malaria forecasting due to their intuitive and robust nature. {ML methods such as long short-term memory (LSTM) networks have also become popular tools for disease prediction due to their ability to model complex, non-linear patterns. However, for this study, which focuses on time series forecasting with structured, moderately sized datasets, we deliberately chose traditional statistical models. This decision was guided by several key factors. Beyond their proven practical performance and methodological clarity, the interpretability aspect of statistical models was a primary consideration. Unlike many ``black-box" ML approaches, the parameters in models like ARIMA correspond directly to interpretable components such as trend, seasonality, and autoregression. This transparency is vital in a public health context, as it allows policymakers to understand the drivers behind a forecast, fostering greater trust and enabling more informed decision-making.} 

The most prominent statistical models can be broadly classified into three methods: Time Series Regression/Generalized Linear Model (GLM) \citep{Hyndman2018, Modeling2019}, Smoothing methods \citep{Hyndman2018}, and Auto-Regressive Integrated Moving Average (ARIMA) \citep{Hyndman2018} methods. Despite previous comparisons among these models, their comparative efficacy in terms of the difference between actual and forecasted malaria cases has not been extensively studied.Time series analysis has emerged as a powerful tool for modeling and forecasting infectious diseases, leveraging historical data to predict future trends.

{A key study by Singh et al.} \citep{Singh2024} {conducted a comprehensive time series analysis of national malaria case trends over the past three decades. By applying ARIMA and Holt's models, the study evaluates the impact of various public health interventions and presents forecasts aligned with India's 2030 malaria elimination goals. While informative at a national level, it does not directly address the need for tools suited for decentralized forecasting at state or district levels, which our study seeks to provide.
At the district level, Chougale et al.} \citep{Chougale2025} {explored malaria forecasting in Mumbai using both classical time series models and machine learning algorithms such as random forests and gradient boosting. Their study demonstrated that no single model uniformly outperformed others, especially when data quality and seasonal variation differed across wards. This reinforces our emphasis on methodological flexibility and model selection based on context and data availability.
Further highlighting the need for region-specific models, a study by Yadav et al.} \citep{Yadav2022} {the forecasting efficacy of eight different statistical models—including Generalized Linear Models (GLM) and various SARIMA configurations—in two semi-arid districts of Gujarat. Their findings identified a specific SARIMA model as the most appropriate for the region, underscoring that optimal model selection is highly dependent on local data characteristics. Although this work fortifies the long-standing utility of statistical methods for sub-national malaria prediction, the lack of a non-linear trend component and a seasonality component separate from ARIMA and SARIMA estimates are aspects that are addressed in our study.
}

In this study, we aim to develop and evaluate multiple time series models for predicting malaria cases across eight Indian states: Chhattisgarh, Jharkhand, Maharashtra, Meghalaya, Mizoram, Odisha, Tripura, and Uttar Pradesh. These states represent high-burden Category 2 and Category 3 states of the stratification by the National Center for Vector-borne Disease Control (NVBDCP) based on the Annual Parasite Index (API). Our objective is to identify the most robust and reliable model for malaria forecasting in these regions. We focus on models that incorporate polynomial trends, seasonal components, and log transformations to enhance prediction accuracy. Furthermore, we have added the results of these models on 18 more states in the supplementary materials.

Further,an interactive R Shiny application is also developed to facilitate the application of these predictive models by researchers and public health officials. This application allows users to input their data, select appropriate models, and visualize predictions and performance metrics, thereby providing a practical tool for real-time malaria forecasting and decision-making support.

By leveraging advanced statistical methods and interactive tools, this research aims to contribute to more effective malaria surveillance and control efforts, ultimately reducing the disease burden in endemic regions.

\section{Methods}
\label{sec2}

\subsection{Time series regression models with autocorrelated errors}
Time series regression models are widely used in various applications to model the relationship between a dependent variable and one or more independent variables over time. One common issue encountered in such models is autocorrelation in the errors, which violates the assumption of independence of residuals in ordinary least squares (OLS) regression. Ignoring autocorrelation can lead to inefficient estimates and misleading inferences \citep{box1976time, tsay2010analysis}.

A linear regression model with autocorrelated errors can be expressed as:
\[
y_t = \mu(t,\beta) + \epsilon_t,
\]
where \( \epsilon_t \) are the zero mean errors that follow an autoregressive moving average model of order \( p, q\) denoted by ARMA(p,q)):
\[
\epsilon_t = \phi_1 \epsilon_{t-1} + \phi_2 \epsilon_{t-2} + \dots + \phi_p \epsilon_{t-p} + u_t + \theta_1 u_{t-1} + \theta_2 u_{t-2} + \dots + \theta_q u_{t-q}
\]
and \( u_t \) is white noise. In the above conditional mean $\mu(t, \beta)$ can be used to capture the trend as a function of time by using suitable basis functions to capture nonlinear trend and it can also include a vector of other explanatory variables $\boldsymbol{x}_t$ which may themselves change over time. For our applications, we have restricted our attention to modeling the trend using orthogonal polynomials of the form $\mu(t, \beta) = \sum_{k=1}^m \beta_k\psi_k(t)$ where $\psi_k(\cdot)$'s denote suitable orthogonal basis functions for $k=1,2,\ldots,m$. More generally, the mean function may be expressed as additive models $\mu(t, \beta) = \sum_{k=1}^m s(x_{kt}, \gamma_k)$, where $s(x,\gamma)$ are some smooth functions of the components of covariate vector $\boldsymbol{x}_t=(x_{1t},\ldots,x_{pt})^\prime$ using B-spilnes or other basis functions. The selection of degree $m$ can be accomplished using cross-validation and information criteria. When residuals from a time series model are autocorrelated, it suggests that the model has not fully captured the dynamics of the data. This autocorrelation can be explicitly modeled to improve the accuracy of the time series model. More generally, the mean function may be expressed as additive models $\mu(t, \beta) = \sum_{k=1}^m s(x_{kt}, \gamma_k)$, where $s(x,\gamma)$ are some smooth functions of the components of covariate vector $\boldsymbol{x}_t=(x_{1t},\ldots,x_{pt})^\prime$ using B-splines or other basis functions \citep{wood2017gam}. However, for the ease of illustration in this paper we consider only polynomial trend without any covariates. But the methodologies described in the paper can be easily extended to more general trend functions (of time) and other potential predictor variables. Next we provide some basic notions of time series modeling which will be used for our subsequent applications.

\subsubsection{Decomposition of Time Series: Trend, Seasonality, and Residuals}

Time series data can often be decomposed into three primary components: trend, seasonality, and residuals. This decomposition is foundational in understanding the underlying patterns within the data and is essential for effective modeling and forecasting. By separating the trend, seasonality, and residual components, we can gain valuable insights into the data's behavior over time. When residuals are autocorrelated, modeling them explicitly through autoregressive models can lead to improved model performance and more accurate forecasts.

\paragraph{Trend Component}

The trend component captures the long-term movement in the data. This is often due to underlying factors such as economic growth, technological advancements, or changes in consumer behavior. Trends can be linear, nonlinear, or even exhibit structural breaks over time \citep{cleveland1990stl}. The trend component can be modeled using various techniques, depending on its complexity. For simple trends, a linear model with polynomial terms in time can be used:
\[
\text{Trend}(t) = \beta_0 + \beta_1 t + \beta_2 t^2 + \dots + \beta_m t^m,
\]
where \( t \) is time and \( \beta_0, \beta_1, \dots, \beta_m \) are the coefficients to be estimated \citep{draper1998applied}. Alternatively, an orthogonal polynomials can be used which would aid the selection of the appropriate degree $m$ \citep{narula1979orthogonal}.

In more complex scenarios, non-parametric methods such as smoothing splines or locally weighted regression (LOESS) can be employed \citep{cleveland1979robust}. These methods do not assume a specific functional form for the trend, allowing for greater flexibility in capturing nonlinear patterns.

\paragraph{Seasonality Component}

Seasonality refers to regular, periodic fluctuations within a time series, typically driven by seasonal factors like the time of year, day of the week, or time of day. Seasonality is characterized by a fixed and known frequency and is assumed to repeat over time \citep{Hyndman2018}. Seasonality can be modeled using dummy variables or Fourier terms, depending on the periodicity and complexity of the seasonal pattern. For instance, a simple seasonal model with monthly data can be written as:
\[
\text{Seasonality}(t) = \gamma_1 D_{1,t} + \gamma_2 D_{2,t} + \dots + \gamma_{11} D_{11,t},
\]
where \( D_{i,t} \) are dummy variables representing the months of the year, and \( \gamma_1, \gamma_2, \dots, \gamma_{11} \) are the seasonal coefficients. Alternatively, Fourier series can be used to model more complex seasonal patterns:
\[
\text{Seasonality}(t) = \sum_{k=1}^K \left[ a_k \cos\left(\frac{2\pi k t}{T}\right) + b_k \sin\left(\frac{2\pi k t}{T}\right) \right],
\]
where \( T \) is the period of the seasonality and \( K \) is the number of harmonics included \citep{Hyndman2018}.

\paragraph{Residual Component}

The residual component, often referred to as the ``remainder" or ``irregular" component, is what remains after removing the trend and seasonality from the original time series. Residuals are expected to be random noise, ideally following a white noise process. However, in practice, residuals often exhibit autocorrelation, indicating the presence of unexplained structure in the data. Once the trend and seasonality have been modeled, the remaining residuals can be examined for autocorrelation. If autocorrelation is present, an autoregressive model can be fitted to the residuals. For example, if the residuals follow an AR(p) process, the model can be expressed as:
\[
\epsilon_t = \sum_{l=1}^p\phi_1 \epsilon_{t-l} + u_t,
\]
where \( \epsilon_t \) are the residuals and \( u_t \) is white noise.

The complete time series model incorporating trend, seasonality, and autocorrelated residuals can thus be written as:
\[
y_t = \text{Trend}(t) + \text{Seasonality}(t) + \epsilon_t,
\]
with \( \epsilon_t \) following an ARIMA process or another suitable autoregressive model \citep{box1976time}. Although Autoregressive Moving Average (ARMA) models are fundamental tools in time series analysis, Hannan and Rissanen (1982) showed that long autoregressive (AR) models can provide consistent estimates for ARMA parameters when the order is large enough and determined using appropriate criteria such as Akaike’s Information Criterion (AIC). So, we are going to restrict our errors process to only AR models.

\subsubsection{Model Evaluation: Adequacy and Accuracy}
Evaluating the adequacy and performance of time series models is a crucial step in the modeling process. This involves both in-sample diagnostics, which assess how well the model fits the historical data, and out-of-sample evaluation, which tests the model's predictive power on unseen or hold out data. A combination of residual analysis and forecast accuracy metrics is typically employed to ensure that the model is both statistically sound and practically useful.

Residual analysis is the first step in time series model checking. Residuals, defined as the difference between the observed and fitted values, should ideally behave like white noise—exhibiting no autocorrelation, having constant variance, and following a normal distribution \citep{box1976time}.

\paragraph{Model Adequacy Based on In-Sample Diagnostics}
We first describe the in-sample diagnostics to check the suitability of various models that we fit to our data sets.Let \( y_t \) is the observed value, \( \hat{y}_t \) is the predicted value based on a fitted model, and \( n \) is the number of observations. We compute the estimated residuals $\hat{\epsilon}_t = y_t - \hat{y}_t$ for $t=1,2,\ldots,n$ and use RMSE and MAPE metrics to evaluate the model adequacy. Key in-sample diagnostics for residuals include:

\begin{itemize}
    \item \textbf{Autocorrelation Function (ACF) of Residuals:} The ACF plot of residuals should show no significant autocorrelation at any lag. Significant autocorrelation suggests that the model has not adequately captured the temporal structure of the data.
    \item \textbf{Ljung-Box Test:} This statistical test is used to assess whether the residuals are independently distributed. A non-significant p-value indicates that there is no evidence of autocorrelation, supporting the model's adequacy.
    \item \textbf{Normality Tests:} Residuals are often assumed to be normally distributed. This can be checked using a Q-Q plot or statistical tests like the Shapiro-Wilk test. Deviations from normality may indicate model misspecification.
\end{itemize}

\paragraph{Forecast Accuracy Based on Out-of-Sample Prediction}

To evaluate the predictive performance of a time series model, it is important to test the model on data not used during the model fitting process. This can be done through a variety of accuracy metrics, with the most common being Root Mean Squared Error (RMSE) and Mean Absolute Percentage Error (MAPE). Suppose we use the data set ${\cal D}_n=\{y_1, y_2, \ldots, y_n\}$ to fit a model and let ${\cal T}_h = \{y_{n+1}, y_{n+2},\ldots,y_{n+h}\}$ denote the test values that are left out for checking forecast accuracy.

RMSE is one of the most widely used metrics for assessing the accuracy of a model’s predictions. It is defined as:
\[
\text{RMSE} = \sqrt{\frac{1}{h} \sum_{t=1}^{h} (y_{n+t} - \hat{y}_{n+t})^2},
\]
RMSE penalizes large errors more heavily than smaller ones due to the squaring of errors, making it sensitive to outliers. A lower RMSE indicates a better fit and more accurate predictions \citep{Hyndman2018}.

MAPE is another popular metric, particularly useful when dealing with data on different scales. It is defined as:
\[
\text{MAPE} = \frac{100}{h} \sum_{t=1}^{h} \left| \frac{y_{n+t} - \hat{y}_{n+t}}{y_{n+t}} \right|,
\]
MAPE expresses the accuracy as a percentage, making it easy to interpret. However, MAPE can be problematic when actual values \( y_{n+t} \) are close to zero, as it can result in very large values or even undefined behavior \citep{armstrong1985long}.

Model checking through residual analysis and out-of-sample prediction evaluation is essential for verifying the adequacy of a time series model. Metrics such as RMSE and MAPE, alongside other diagnostic tools, provide a comprehensive understanding of model performance. By rigorously assessing both in-sample fit and out-of-sample accuracy, one can ensure that the model is both statistically valid and practically useful.

\paragraph{Rolling Forecast}\label{rolling}

A rolling forecast approach is used to evaluate the models' performance over time. The data is split into training and testing sets in a rolling manner, and forecasts are made for a specified number of steps ahead (e.g., 3 steps). MAPE is calculated for each rolling forecast to assess the prediction accuracy of the models over different time windows. MAPE values from the rolling forecasts are plotted to visualize the models' prediction accuracy over time. 

\section{Methodology}

\subsection{Data and preprocessing}
The state-level malaria case counts are obtained from the Monthly Malaria Information System\,(MMIS) from the National Centre Vector Borne Diseases Control\,(NCVBDC).{At the time of the analysis, the most recent validated data available was up to December 2023.} The data contains monthly malaria case counts of various states from January 2020 to December 2023. Based on the annual parasite index\,(API), Indian states and union territories are categorised into three strata, Category III consisting of the highest-burden states.
It is read from CSV file and converted to a \texttt{Date} type and sorted by date to ensure chronological order. Month information is extracted and converted to dummy variables to account for seasonality in the models. A numeric time column is created to represent the date, which is used in the polynomial trend models.

\subsubsection{Trend Estimation, Seasonal Adjustments and Residual}

The general modeling framework we have utilized is:
\begin{align}    
Y_t &= \mu_t + \epsilon_t, \nonumber\\
\mu_t &= \beta_0 + \sum_{k=1}^m \beta_kt^k \nonumber
\end{align}
where $\beta_0$ is the seasonal component and $t, t^2, t^3 \dots t^m$ is a sequence of $m$ basis functions which capture the nonlinear trend component. The trend component and seasonal component of the case counts were modeled using four variants of these regression models, and a fifth model which improves upon the best performing model of the former four models by reducing the autocorrelation within the residuals. The approach we have followed is Algorithm \ref{algo}.

\begin{algorithm}
\caption{Malaria Cases Forecasting with ARIMA Error Correction}
\begin{algorithmic}[1]
    \Require Monthly malaria cases vector $x$
    \Ensure $x$ has at least $36$ data points
    \State Pre-process $x$
    \State Perform 3 month out-of-sample rolling forecasts as detailed in section \ref{rolling} for Models \ref{eqn: 1}, \ref{eqn: 2}, \ref{eqn: 3} and \ref{eqn: 4} and select Model $M$ with lowest average rolling forecast MAPE
    \State Perform ARIMA modeling on the residuals of $M$ to get Model \ref{eqn: 5} and get out-of-sample rolling forecasts
    \State Compare the average rolling forecast MAPE of Model \ref{eqn: 5} with $M$ and select the best performing model
    \State Obtain in-sample fits for Models \ref{eqn: 1}, \ref{eqn: 2}, \ref{eqn: 3}, \ref{eqn: 4} and \ref{eqn: 5} on $x$, and calculate metrics
    \State \textbf{return} $M$, plots and metrics
\end{algorithmic}
\label{algo}
\end{algorithm}

\paragraph{Model 1 (Poly + Season)}
\begin{align} 
\label{eqn: 1}
y_t &= \alpha_{s_{t}} + \beta_1 t + \beta_2 t^2 + \epsilon_t, \nonumber\\
\hat{y}_t &= \hat{\alpha}_{s_{t}} + \hat{\beta}_1 t + \hat{\beta}_2 t^2. 
\end{align}
This model incorporates a quadratic polynomial trend and seasonal dummy variables to account for periodic patterns in the case counts. The raw (unnormalized) case counts are used as the response variable, ensuring direct interpretability. The seasonal component (\(s_t\)) captures recurring fluctuations, while the polynomial terms (\(t, t^2\)) model the overall trend.


\paragraph{Model 2 (Log Transformed)}
\begin{align} 
\label{eqn: 2}
\log y_t &= \tilde{\alpha}_{s_{t}} + \hat{\beta}_1 t + \hat{\beta}_2 t^2 + \epsilon_t, \nonumber\\ 
\hat{y_t} &= \exp(\tilde{\alpha}_{s_{t}} + \hat{\beta}_1 t + \hat{\beta}_2 t^2 + 0.5 \sigma^2).  
\end{align}
This model uses log counts with polynomial trend and seasonal dummies. When transforming back from log counts, the variance of residuals is adjusted using the mean of a log-normal distribution, $E[Y_t] = \exp\{\mu_t + 0.5\sigma^2\}$ \citep{logcorrection} where $\log Y_t \sim N (\mu_t, \sigma^2)$, with $\mu_t = s_t + \beta_1 t + \beta_2 t^2$.

\paragraph{Model 3 (Lag + Trend)}
\begin{align}
\label{eqn: 3}
y_t &= y_{t-12} + \beta_0 + \beta_1 t + \beta_2 t^2 + \epsilon_t, \nonumber\\
\hat{y_t} &= y_{t-12} + \hat{\beta}_1 t  + \hat{\beta}_2 t^2.
\end{align}
This approach incorporates a lag-12 differencing term (\(y_{t-12}\)) to directly model the seasonal effect by referencing values from the same period in the previous year. The quadratic polynomial trend accounts for long-term changes, while the differencing effectively captures and removes seasonality in the unnormalized case counts.

\paragraph{Model 4 (Lag + Poly Trend)}
\begin{align}
\label{eqn: 4}
y_t &= \gamma y_{t-12} + \beta_1 t + \beta_2 t^2 + \epsilon_t, \nonumber\\
\hat{y_t} &= \hat{\gamma} y_{t-12} + \hat{\beta}_1 t + \hat{\beta}_2 t^2 + \epsilon_t. 
\end{align}
This model builds on Model \ref{eqn: 3} by incorporating a multiplicative factor (\(\gamma\)) for the lag-12 term, allowing for an autoregressive seasonal component. This dynamic seasonal adjustment is particularly useful for capturing subtle interactions between seasonality and trends. Polynomial terms model the trend, while the lag-12 term adjusts for recurring annual variations.

For all models, residuals (\(\epsilon_t\)) are analyzed using autoregressive (AR) or autoregressive integrated moving average (ARIMA) models to account for any remaining temporal dependencies. These residual models enhance forecasting accuracy by capturing patterns unexplained by the main regression model.

\paragraph{Model 5 (ARIMA)}

After identifying the best-fitting model based on performance metrics such as MAPE and RMSE, the residuals from this model will be used to fit an ARIMA model.

\begin{equation}
\label{eqn: 5}
\epsilon_t^{\text{best}} = \phi_0 + \phi_1 \epsilon_{t-1} + \phi_2 \epsilon_{t-2} + \cdots + \phi_p \epsilon_{t-p} + \theta_1 e_{t-1} + \theta_2 e_{t-2} + \cdots + \theta_q e_{t-q} + e_t.
\end{equation}

The presence of autocorrelation in the residuals, denoted as \(\epsilon_t^{\text{best}}\), indicates that temporal dependencies remain unexplained by the regression model. Fitting an ARIMA model to these residuals helps capture and model these dependencies, further improving the overall predictive accuracy.


\section{Results and Discussion}

The statistical analysis involves fitting multiple regression models to the time series data, including polynomial trends and seasonal components, and evaluating their performance using residual analysis and forecasting accuracy metrics. The rolling forecast approach provides a robust assessment of the models' predictive performance over time, which is crucial for understanding their reliability in real-world applications. This comprehensive analysis helps in selecting the best-performing model for forecasting and provides insights into the underlying patterns and trends in the data.

\begin{table}[ht]
\centering
\caption{In-sample model performance metrics for different states.}
\label{tab: results}
\begin{tabular}{lrrlrr}
\toprule
\textbf{State/Model} & \textbf{RMSE} & \textbf{MAPE} & \textbf{State/Model} & \textbf{RMSE} & \textbf{MAPE} \\
\cmidrule(r){1-3} \cmidrule(l){4-6}
\textbf{Chhattisgarh} & & & \textbf{Mizoram} & & \\
Poly + Season & 5.0330 & 0.0443 & Poly + Season & 6.3836 & 0.1182 \\
Log Transformed & 5.2502 & 0.0462 & Log Transformed & 7.6807 & 0.1104 \\
Lag + Trend & 7.6914 & 0.0633 & Lag + Trend & 10.3390 & 0.1631 \\
Lag + Poly Trend & 7.4476 & 0.0610 & Lag + Poly Trend & 10.3389 & 0.1632 \\
ARIMA & 7.4476 & 0.0610 & ARIMA & 5.9463 & 0.1066 \\
\cmidrule(r){1-3} \cmidrule(l){4-6}
\textbf{Jharkhand} & & & \textbf{Odisha} & & \\
Poly + Season & 18.5048 & 0.1754 & Poly + Season & 7.1806 & 0.0624 \\
Log Transformed & 19.1094 & 0.1585 & Log Transformed & 7.6663 & 0.0619 \\
Lag + Trend & 32.6291 & 0.2769 & Lag + Trend & 10.1003 & 0.0922 \\
Lag + Poly Trend & 25.5615 & 0.2730 & Lag + Poly Trend & 10.0956 & 0.0921 \\
ARIMA & 18.5048 & 0.1754 & ARIMA & 7.1480 & 0.0558 \\
\cmidrule(r){1-3} \cmidrule(l){4-6}
\textbf{Maharashtra} & & & \textbf{Tripura} & & \\
Poly + Season & 10.4354 & 0.1440 & Poly + Season & 13.0241 & 0.2478 \\
Log Transformed & 9.4180 & 0.1136 & Log Transformed & 13.6504 & 0.2487 \\
Lag + Trend & 17.5058 & 0.2069 & Lag + Trend & 24.6156 & 0.4655 \\
Lag + Poly Trend & 17.0061 & 0.2106 & Lag + Poly Trend & 23.5940 & 0.4704 \\
ARIMA & 8.8177 & 0.0983 & ARIMA & 12.2266 & 0.1515 \\
\cmidrule(r){1-3} \cmidrule(l){4-6}
\textbf{Meghalaya} & & & \textbf{Uttar Pradesh} & & \\
Poly + Season & 4.2975 & 0.2840 & Poly + Season & 20.2153 & 0.3457 \\
Log Transformed & 3.3672 & 0.1702 & Log Transformed & 12.0958 & 0.1777 \\
Lag + Trend & 5.4883 & 0.3890 & Lag + Trend & 27.6193 & 0.5838 \\
Lag + Poly Trend & 5.0019 & 0.3170 & Lag + Poly Trend & 18.0696 & 0.4331 \\
ARIMA & 2.4789 & 0.1448 & ARIMA & 9.8979 & 0.1557 \\
\bottomrule
\end{tabular}
\end{table}

Table \ref{tab: results} presents the performance of five forecasting models—Poly + Season, Log Transformed, Lag + Trend, Lag + Poly Trend, and ARIMA—across eight Indian states using Root Mean Squared Error (RMSE) and Mean Absolute Percentage Error (MAPE) as performance metrics. Overall, the Log Transformed and ARIMA models consistently delivered strong results across multiple states. In Odisha, the Log Transformed model achieved the lowest MAPE (0.0619), while ARIMA performed best in terms of RMSE (7.1480). A similar pattern emerged in Jharkhand, where the Log Transformed model showed a lower MAPE (0.1585) compared to ARIMA, which had the lowest RMSE (18.5048). For Chhattisgarh, the Poly + Season model demonstrated superior performance with the lowest RMSE (5.0330) and MAPE (0.0443), indicating its robustness. In Maharashtra, ARIMA recorded the lowest RMSE (8.8177), while the Log Transformed model had the lowest MAPE (0.1136). In Tripura, the Log Transformed model excelled in percentage error minimization (MAPE of 0.2487), while ARIMA showed a lower RMSE (12.2266). In Meghalaya, ARIMA was the best-performing model across both metrics, with the lowest RMSE (2.4789) and MAPE (0.1448). Similarly, for Uttar Pradesh, the Log Transformed model provided the best accuracy in terms of MAPE (0.1777), while ARIMA achieved a lower RMSE (9.8979). In Mizoram, ARIMA had the lowest RMSE (5.9463), though the Log Transformed model once again minimized percentage errors with a MAPE of 0.1104. Across most states, the Log Transformed model generally performed better in minimizing MAPE, while ARIMA showed relatively better performance in reducing RMSE.

\begin{figure}[ht]
\centering
\includegraphics[width=0.496\textwidth]{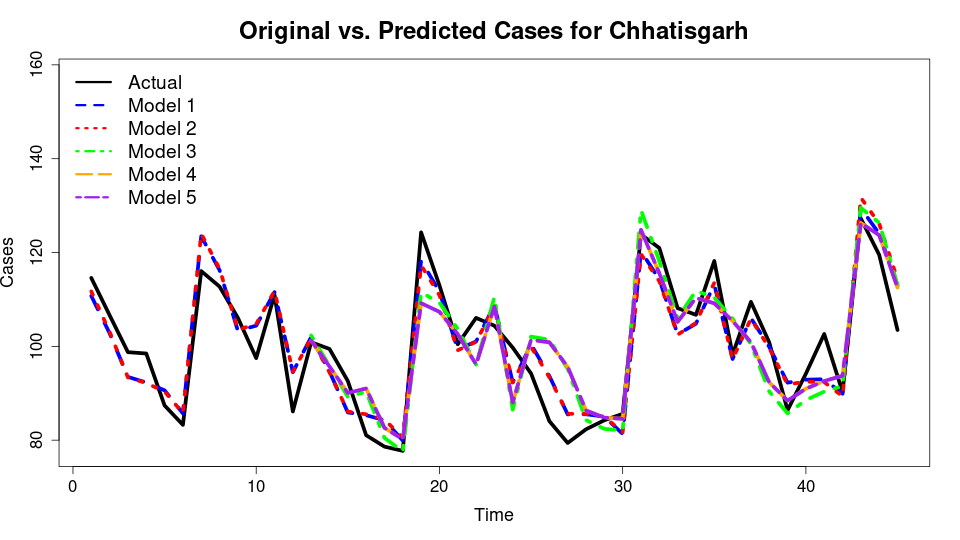}
\includegraphics[width=0.496\textwidth]{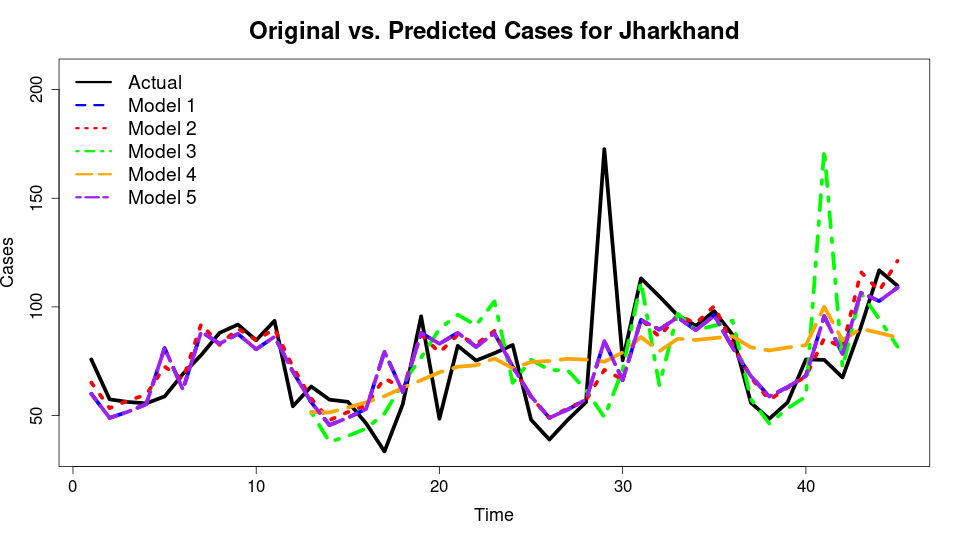}
\includegraphics[width=0.496\textwidth]{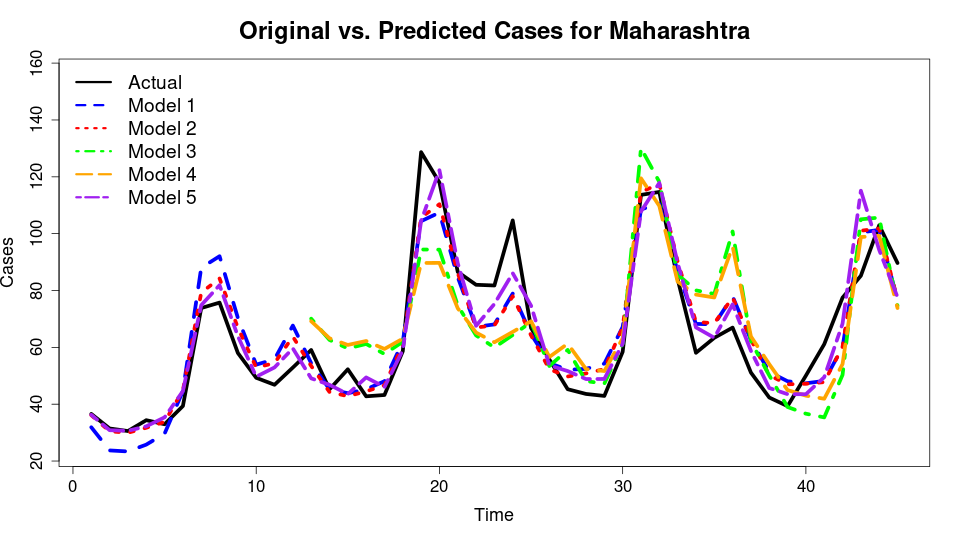}
\includegraphics[width=0.496\textwidth]{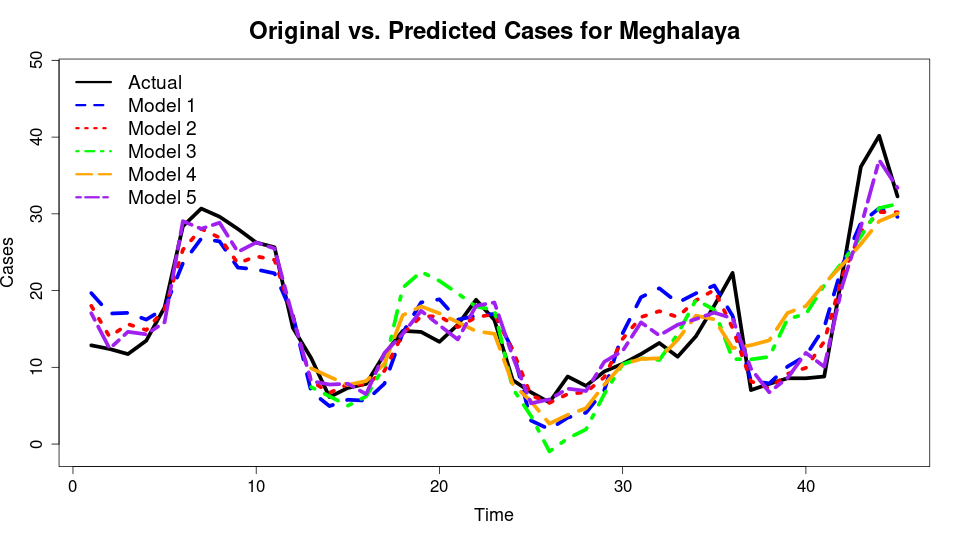}
\includegraphics[width=0.496\textwidth]{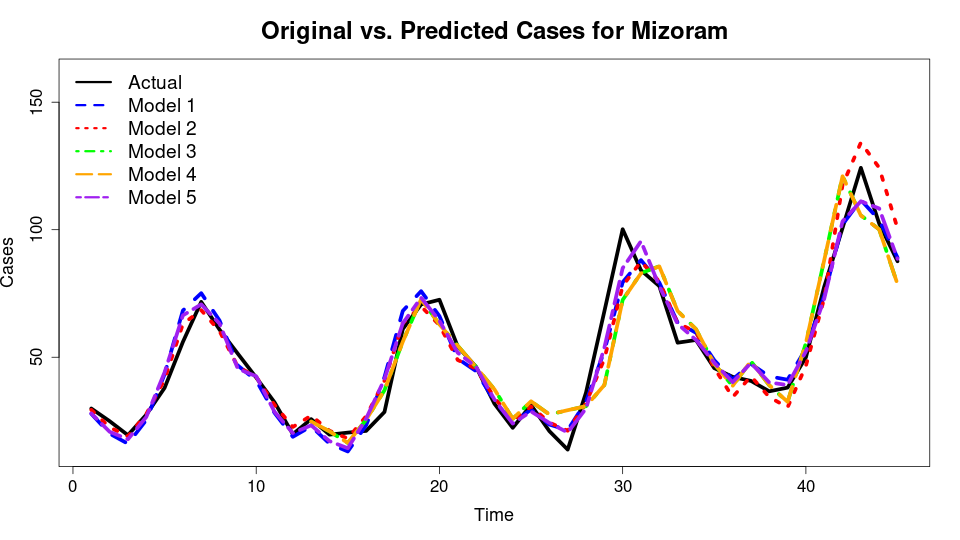}
\includegraphics[width=0.496\textwidth]{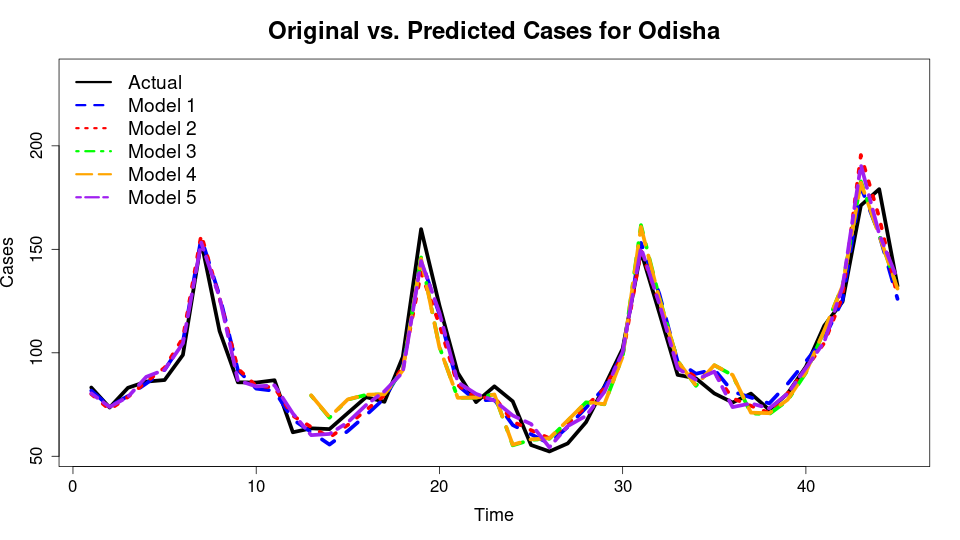}
\includegraphics[width=0.496\textwidth]{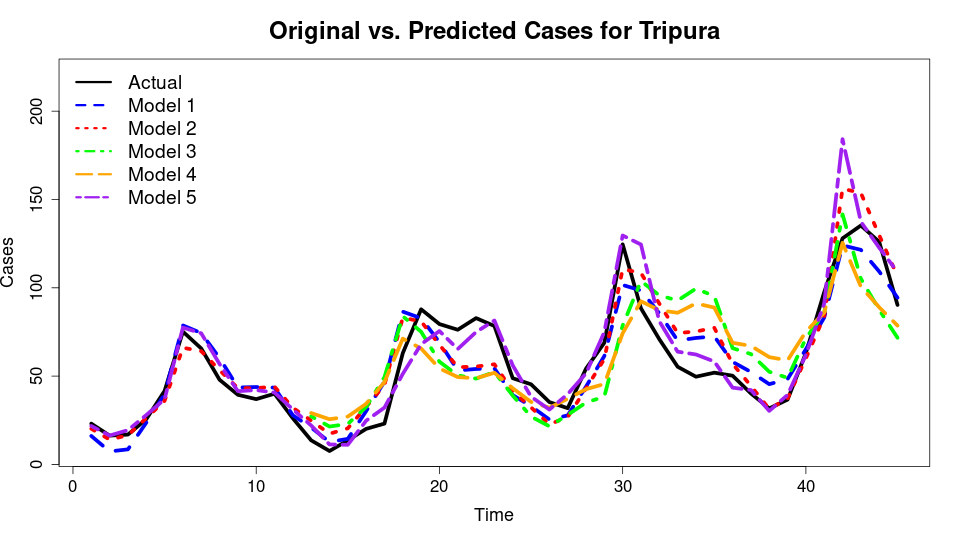}
\includegraphics[width=0.496\textwidth]{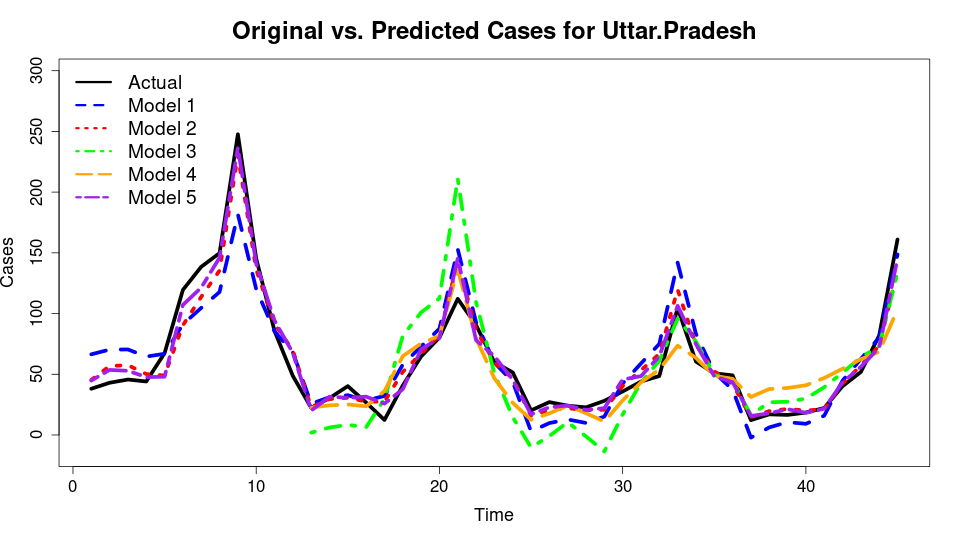}
\caption{Plots illustrating original vs in-sample predicted cases for all models in different states.}
\label{fig:plots}
\end{figure}
The figure \ref{fig:plots} presents plots of original versus predicted cases for various forecasting models across eight states: Chhattisgarh, Jharkhand, Maharashtra, Meghalaya, Mizoram, Odisha, Tripura, and Uttar Pradesh. In Odisha and Maharashtra, the Log Transformed model closely matches the actual data, indicating high accuracy. For Jharkhand, the models exhibit varying degrees of accuracy, with noticeable divergence during peak periods. In Chhattisgarh, the Lag + Poly Trend model captures the data fluctuations effectively. The Log Transformed model performs best in Uttar Pradesh, accurately predicting the large fluctuations. In Tripura, the Poly + Season model aligns closely with the actual trends. These observations highlight that while some models consistently perform well, the optimal model varies by state, emphasizing the importance of tailored model selection for accurate forecasting.

\begin{figure}[ht]
\centering
\includegraphics[width=0.496\textwidth]{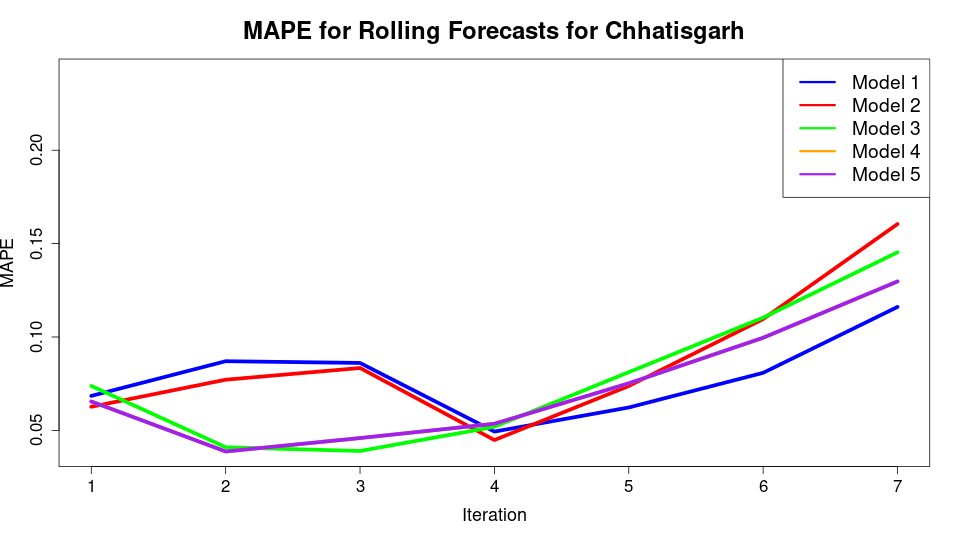}
\includegraphics[width=0.496\textwidth]{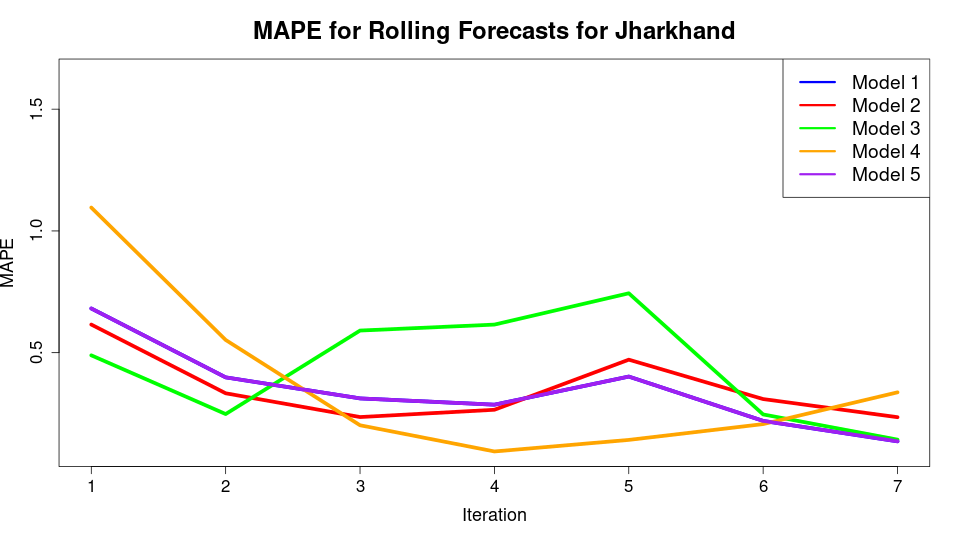}
\includegraphics[width=0.496\textwidth]{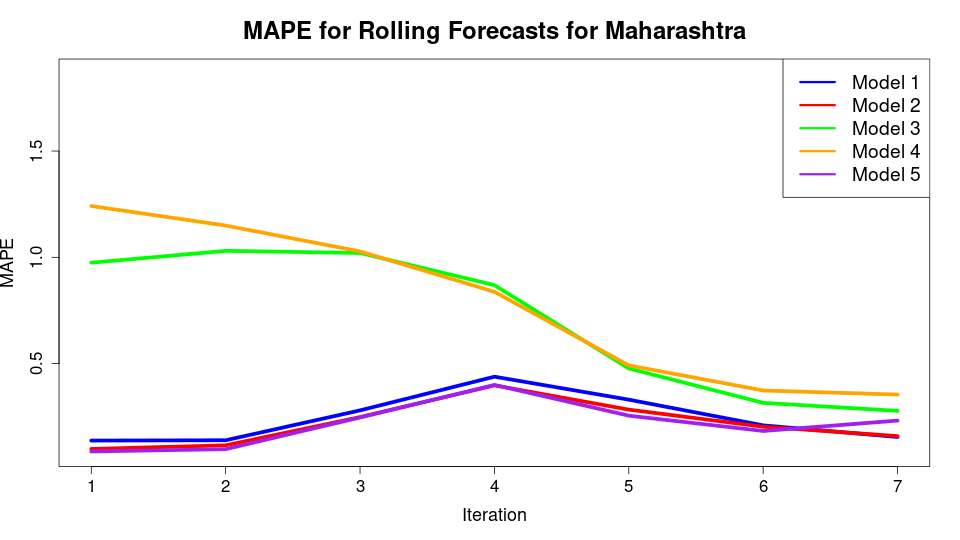}
\includegraphics[width=0.496\textwidth]{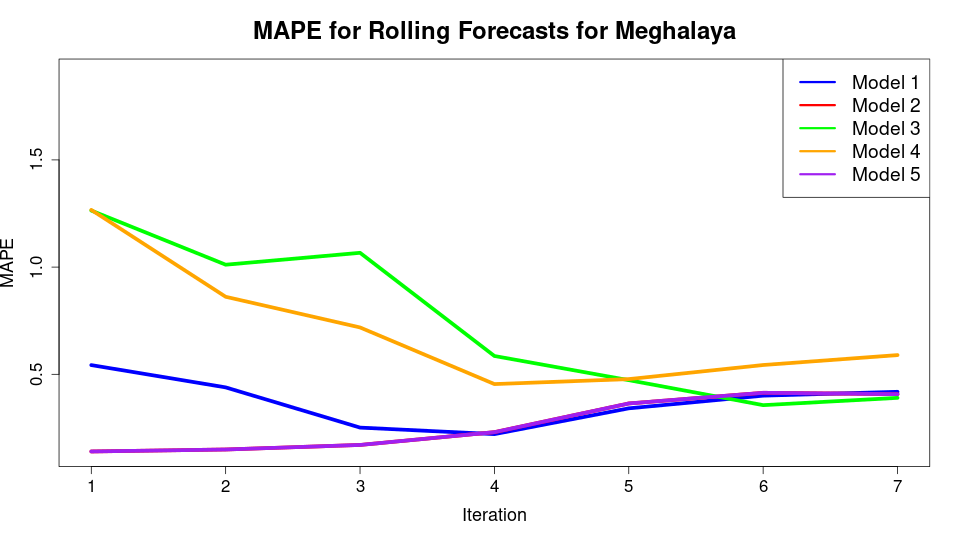}
\includegraphics[width=0.496\textwidth]{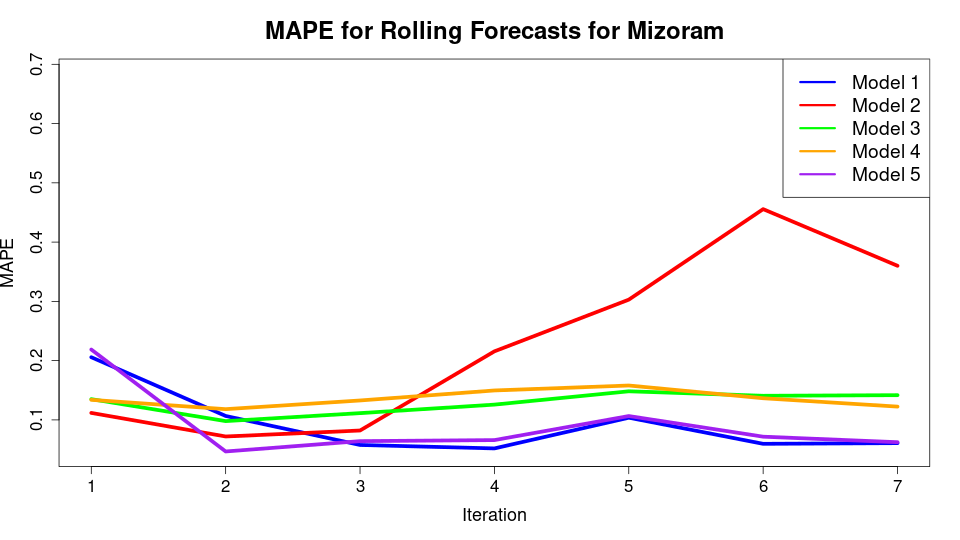}
\includegraphics[width=0.496\textwidth]{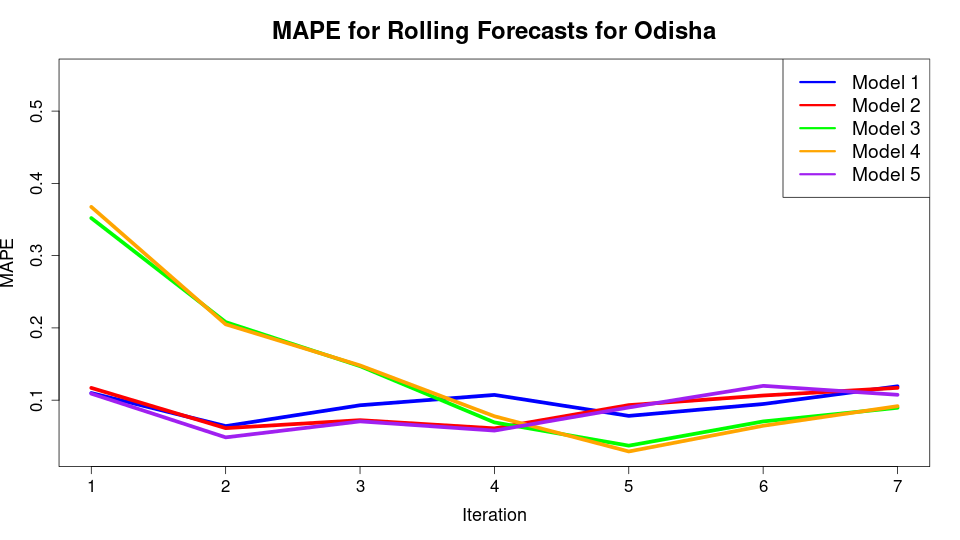}
\includegraphics[width=0.496\textwidth]{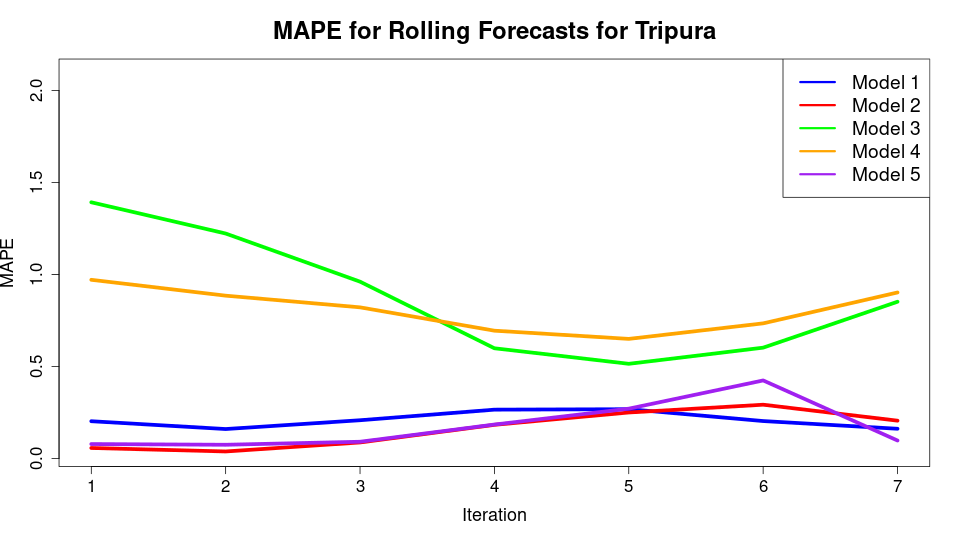}
\includegraphics[width=0.496\textwidth]{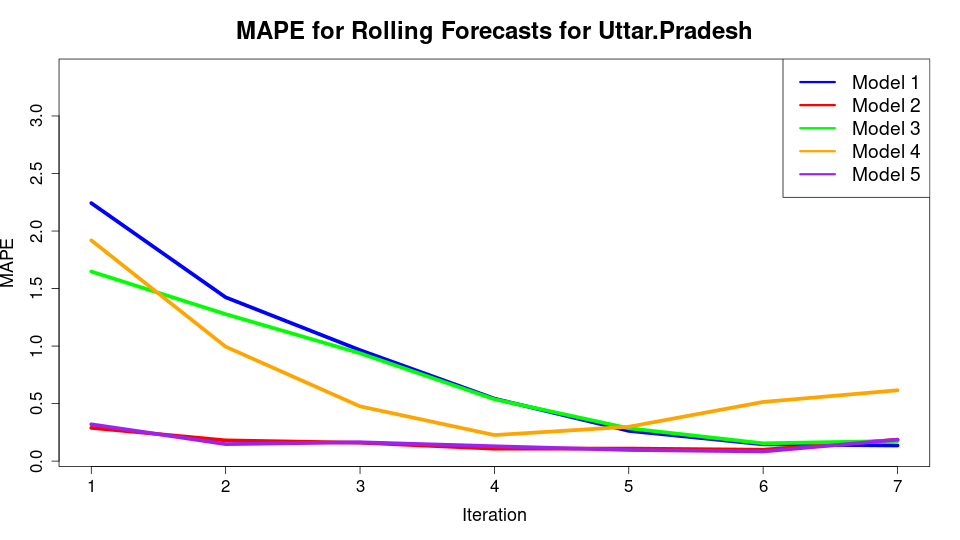}
\caption{Plots illustrating model out of sample MAPE}
\label{fig:plotts}
\end{figure}
The figure \ref{fig:plotts} shows the Mean Absolute Percentage Error (MAPE) for rolling forecasts across eight states using five models. The Log Transformed model consistently delivers the lowest MAPE values, particularly excelling in Odisha, Maharashtra, and Uttar Pradesh. The Poly + Season model performs best in Tripura, while the Lag + Trend model shows strong results in Jharkhand. The ARIMA model improves over iterations but generally starts with higher MAPE values. These results highlight the need for tailored model selection to optimize forecasting accuracy for each state's unique data characteristics.

Table \ref{tab:results} presents the out-of-sample Mean Absolute Percentage Error (MAPE) performance evaluation for eight states—Chhattisgarh, Jharkhand, Maharashtra, Meghalaya, Mizoram, Odisha, Tripura, and Uttar Pradesh. The performance metrics were derived using five different models, and the MAPEs were calculated for each model, namely MAPE1, MAPE2, MAPE3, MAPE4, and MAPE5. In this study, MAPE1 represents the model trained on 36 data points and tested on 8, MAPE2 trained on 37 data points and tested on 7, MAPE3 trained on 38 data points and tested on 6, MAPE4 trained on 39 data points and tested on 5, and MAPE5 trained on 40 data points and tested on 4. This allows for a progressive evaluation of the model's performance as the amount of training data increases while the test set decreases.

For Chhattisgarh, MAPE5 and MAPE4 exhibit the lowest MAPE value of 0.0727, indicating higher accuracy for these models. In Jharkhand, both MAPE1 and MAPE5 models provide the best performance with a MAPE value of 0.3484, highlighting that these models are more reliable for this state. For Maharashtra, MAPE5 demonstrates superior accuracy with the lowest MAPE value of 0.2132, followed closely by MAPE2. Similarly, in Meghalaya, MAPE5 and MAPE2 perform the best, with MAPE5 achieving a MAPE of 0.2675. In Mizoram, MAPE5 shows the highest accuracy with a MAPE value of 0.0960, while in Odisha, MAPE1 stands out with a MAPE of 0.0956, indicating it as the most accurate model for this state. For Tripura, MAPE2 performs the best, yielding the lowest MAPE value of 0.1596. In Uttar Pradesh, MAPE5 exhibits the best performance, achieving the lowest MAPE of 0.1589, demonstrating its robustness for this state's data. MAPE5 generally shows the best performance across multiple states, but the optimal model varies depending on the specific characteristics of each state’s data. MAPE5 excels in Maharashtra, Meghalaya, Mizoram, and Uttar Pradesh, while MAPE2 performs well in Tripura, and MAPE1 excels in Odisha. These findings emphasize the importance of selecting forecasting models that are tailored to the unique data characteristics of each state to achieve the highest prediction accuracy.

\begin{table}[ht]
\centering
\begin{adjustbox}{max width=\textwidth}
\begin{tabular}{lrrrrrrr}
\toprule
\textbf{State/Model} & \textbf{MAPE1} & \textbf{MAPE2} & \textbf{MAPE3} & \textbf{MAPE4} & \textbf{MAPE5} \\
\midrule
\textbf{Chhattisgarh} & 0.0781 & 0.0870 & 0.0774 & 0.0727 & 0.0727 \\
\textbf{Jharkhand} & 0.3484 & 0.3806 & 0.4719 & 0.4220 & 0.3484 \\
\textbf{Maharashtra} & 0.2419 & 0.2144 & 0.8261 & 0.8954 & 0.2132 \\
\textbf{Meghalaya} & 0.3731 & 0.2675 & 0.7351 & 0.7306 & 0.2675 \\
\textbf{Mizoram} & 0.1060 & 0.2556 & 0.1338 & 0.1306 & 0.0960 \\
\textbf{Odisha} & 0.0956 & 0.1049 & 0.1452 & 0.1476 & 0.0995 \\
\textbf{Tripura} & 0.2103 & 0.1596 & 0.8576 & 0.7919 & 0.1922 \\
\textbf{Uttar Pradesh} & 0.9514 & 0.1747 & 0.8174 & 0.6754 & 0.1589 \\
\bottomrule
\end{tabular}
\end{adjustbox}
\caption{Out of sample MAPE averages for each state}
\label{tab:results}
\end{table}

The Table \ref{tab:lowest_mape} presents the out-of-sample Mean Absolute Percentage Error (MAPE) for five forecasting models—Poly + Season, Log Transformed, Lag + Trend, Lag + Poly Trend, and ARIMA—applied across eight states: Chhattisgarh, Jharkhand, Maharashtra, Meghalaya, Mizoram, Odisha, Tripura, and Uttar Pradesh. MAPE measures the accuracy of the models, with lower values indicating better performance.

\begin{table}[ht]
\centering
\begin{adjustbox}{max width=\textwidth}
\begin{tabular}{lrr}
\toprule
\textbf{State} & \textbf{Lowest MAPE Model} & \textbf{MAPE Value} \\
\midrule
\textbf{Chhattisgarh} & Lag + Poly Trend & 0.0727 \\
\textbf{Jharkhand} & Lag + Trend & 0.3806 \\
\textbf{Maharashtra} & Log Transformed & 0.2144 \\
\textbf{Meghalaya} & Log Transformed & 0.2675 \\
\textbf{Mizoram} & Poly + Season & 0.1060 \\
\textbf{Odisha} & Log Transformed & 0.1049 \\
\textbf{Tripura} & Poly + Season & 0.1596 \\
\textbf{Uttar Pradesh} & Log Transformed & 0.1747 \\
\bottomrule
\end{tabular}
\end{adjustbox}
\caption{Best Performing model for each state}
\label{tab:lowest_mape}
\end{table}

The Lag + Poly Trend model performs best in Chhattisgarh, with a MAPE of 0.0727, indicating it as the most reliable model for this state. In Jharkhand, the Lag + Trend model excels with a MAPE of 0.3806. For Tripura and Mizoram, the Poly + Season model yields the lowest MAPE values, achieving 0.1596 and 0.1060, respectively. Despite the Log Transformed model generally providing the most accurate predictions across multiple states, the performance of other models like Lag + Trend, Lag + Poly Trend, and Poly + Season underscores the need for region-specific approaches. For example, in Maharashtra, Meghalaya, Odisha, and Uttar Pradesh, the Log Transformed model stands out, while Jharkhand and Chhattisgarh benefit from the Lag + Trend and Lag + Poly Trend models, respectively. Similarly, the Poly + Season model performs best in Tripura and Mizoram. These observations highlight the critical need for tailored model selection to accommodate the unique characteristics of each state’s data. However, this variability in model performance poses a challenge for practitioners and policymakers.

Efficiently applying the appropriate model for a specific region requires expertise in statistical modeling and a deep understanding of the data. To address this gap, we developed an interactive R Shiny application that simplifies the process by enabling users to easily input data, explore various modeling approaches, and visualize predictions. This tool bridges the gap between sophisticated statistical techniques and practical decision-making by providing an accessible platform for malaria surveillance and forecasting. It empowers researchers and public health officials to make informed, data-driven decisions without needing extensive expertise in time series modeling.

\section{R Shiny web tool}

\subsection{Overview}
We developed an interactive R Shiny application to predict malaria cases using various statistical models. This application simplifies the analysis of malaria data by providing visual tools that assess the performance of different models. It is designed to be user-friendly, catering to both researchers and public health officials. Users can easily input data, select appropriate models, and interpret results effectively, making the tool a valuable asset in malaria surveillance and decision-making.

\subsection{Features}

The application begins by allowing users to upload a CSV file containing malaria case data. The uploaded data is automatically processed and formatted for analysis, ensuring it is ready for model fitting. The app offers flexibility in model selection, including options such as Linear Regression (lm), Robust Linear Regression (rlm), Polynomial + Seasonal, Log Polynomial + Seasonal, Lag 12, Lag + Polynomial Trend, and ARIMA models. This range of models enables users to explore different approaches and identify the most suitable one for their data.

\begin{figure}[ht]
    \centering
    \includegraphics[width=0.8\textwidth]{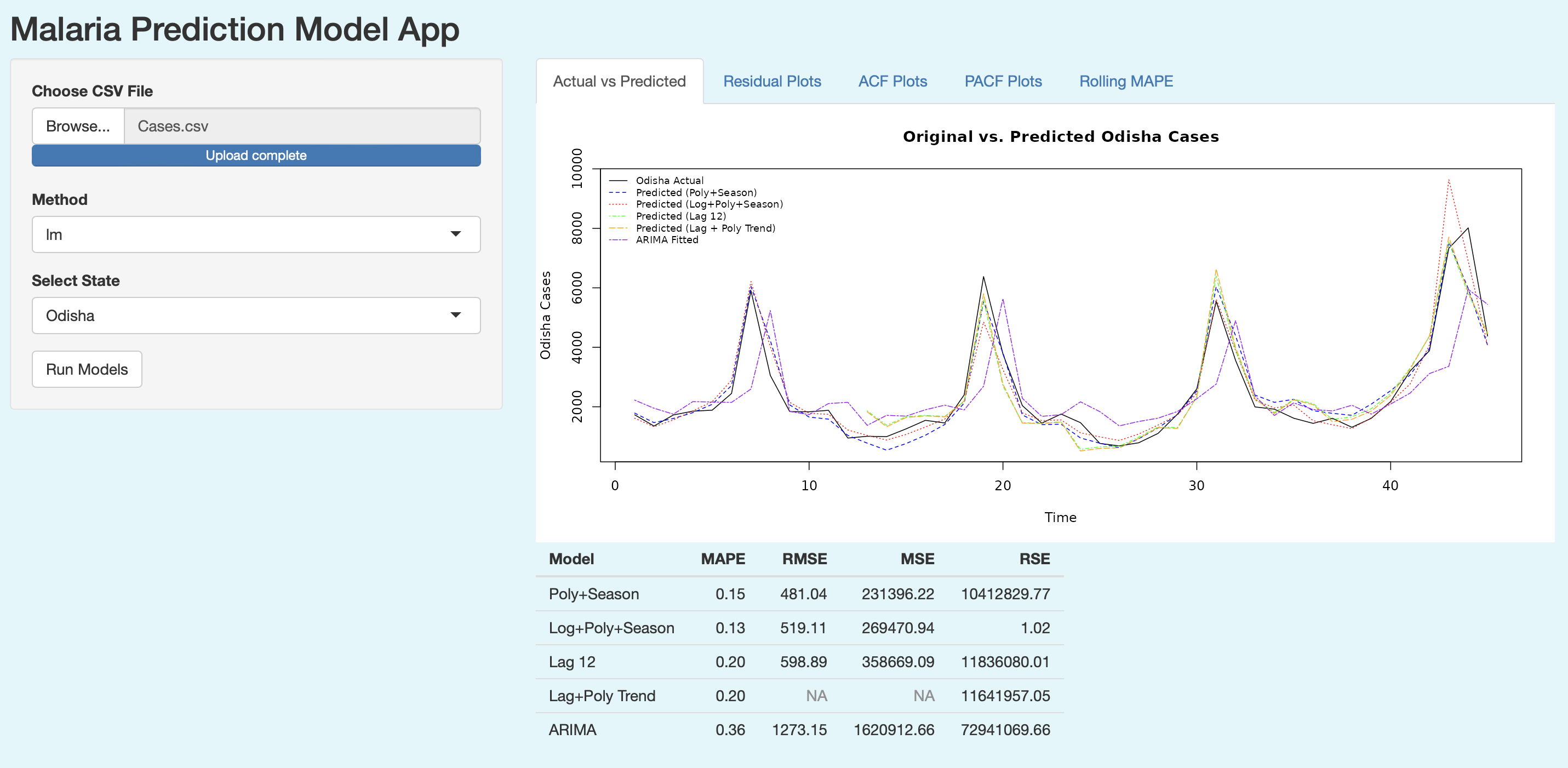}
    \caption{Malaria Prediction Model App Interface}
    \label{fig:shiny_app}
\end{figure}

For regional analysis, the app includes a feature that allows users to select specific states from the dataset. Once a state is selected and a model is chosen, the application generates visual outputs that compare actual versus predicted cases. These visualizations include not only the core comparison of predictions but also diagnostic plots such as residual plots, Autocorrelation Function (ACF) plots, and Partial Autocorrelation Function (PACF) plots. To provide a comprehensive assessment of model performance, the application also calculates key performance metrics, including Mean Absolute Percentage Error (MAPE), Root Mean Squared Error (RMSE), Mean Squared Error (MSE), and Residual Standard Error (RSE).

A unique feature of the application is its rolling forecast analysis. This feature assesses the predictive accuracy of models over time by training them on expanding datasets and making predictions for fixed future periods. The rolling MAPE is plotted over time, offering users a clear view of how model accuracy evolves. This functionality is especially useful in dynamic environments where prediction reliability is critical.

The application's interface is designed for ease of use, with clearly labeled inputs and outputs that guide users through the process. Users can navigate between different tabs to view plots, performance metrics, and rolling forecast results, ensuring a seamless and intuitive experience.

\subsection{Application Workflow}
The process starts with data upload and preparation. Users provide a CSV file containing malaria case data, including a ``Date" column and state-specific case counts. The application automatically preprocesses this data, converting the ``Date" column to the correct format and sorting the data chronologically. Following this, users can select a state and a modeling method. The application then fits multiple models to the data, utilizing methods such as Polynomial + Seasonal, Log Polynomial + Seasonal, Lag 12, Lag + Polynomial Trend, and ARIMA.

Once the models are fitted, the application generates predictions and visualizes these alongside the actual case counts. Additional diagnostic plots, including residuals, ACF, and PACF, are provided to help users diagnose model performance and identify potential issues. The app also calculates and presents various performance metrics, allowing users to compare models in terms of accuracy and robustness.

The rolling forecast analysis further enhances the app's utility by providing ongoing assessments of model accuracy over time. Users can view how the models perform in different time windows, giving them valuable insights into their reliability. The interface is intuitive and designed to facilitate smooth interaction, allowing users to easily switch between different tabs and access all the necessary information.

{While our R Shiny application is designed to accommodate data at various temporal resolutions—such as hourly, weekly, or monthly-scaling the tool to district-level forecasting introduces several practical limitations such as limited data availability    
data quality and noise, heterogeneity in reporting systems, etc. To mitigate these challenges, the R Shiny tool includes preprocessing options that allow users to aggregate high-resolution data (e.g., hourly counts) into coarser temporal units (e.g., daily or weekly averages). This aggregation can help stabilize variance and improve model fit when high-frequency noise is dominant. However, users should be mindful that reducing temporal granularity may lead to a loss of short-term dynamics, and such trade-offs should be considered based on the intended use case.}

\section{Conclusion}

This study successfully developed and evaluated several time series models for predicting malaria cases across eight Indian states: Chhattisgarh, Jharkhand, Maharashtra, Meghalaya, Mizoram, Odisha, Tripura, and Uttar Pradesh. Among the models tested, the Log Transformed model consistently demonstrated superior performance in most states, including Maharashtra, Odisha, Meghalaya, and Uttar Pradesh, as evidenced by lower Mean Absolute Percentage Error (MAPE).

The analysis emphasized the importance of selecting appropriate data transformations and seasonal components to enhance the accuracy of predictions. For states such as Chhattisgarh and Jharkhand, Lag + Poly Trend and Lag + Trend models provided better accuracy, showcasing that the optimal model can vary depending on the unique characteristics of each region's data. The Poly + Season model also performed notably well in states like Mizoram and Tripura, indicating its effectiveness in capturing seasonal trends for those areas.

While the Log Transformed model generally outperformed others, the study highlights that a one-size-fits-all approach may not be suitable for malaria forecasting across different states. The variability in model performance across states underlines the need for region-specific model selection. In comparison, the ARIMA model, though widely applied in time series forecasting, showed relatively higher error rates, confirming its limitations for this particular application.

Additionally, the development of an interactive R Shiny application enhances the practical utility of these findings. This application allows users to upload data, choose from multiple modeling approaches, and visualize predictions with performance metrics. By offering a user-friendly interface, the app supports real-time malaria forecasting, enabling public health officials and researchers to make data-driven decisions for disease surveillance and intervention planning.

{The results suggest that tailored model selection is essential for optimal accuracy. The R Shiny application developed as part of this research serves as a valuable tool for ongoing malaria surveillance and forecasting efforts using univariate time series models. Rolling forecast validation was implemented to mitigate the overfitting of models, but continuous validation with incoming data remains essential to ensure sustained model performance. The modeling approach in this study is currently limited to state-level aggregation; extending this approach to finer geographic units such as districts would require addressing challenges like data sparsity, inconsistent reporting, and variability in surveillance infrastructure. Further, future research could explore potentially important external predictors such as climatic variables, intervention coverage, and socioeconomic conditions. Incorporating such covariates into multivariate models could enhance both predictive accuracy and explanatory power.}

\section{Funding}

This research received no dedicated or external funding.

\bibliography{references}

@book{tsay2010analysis,
  title={Analysis of Financial Time Series},
  author={Ruey S. Tsay},
  series={Wiley Series in Probability and Statistics},
  year={2010},
  publisher={John Wiley \& Sons}
}

@article{narula1979orthogonal,
 ISSN = {03067734, 17515823},
 doi = {10.2307/1403204},
 author = {Sabhash C. Narula},
 journal = {International Statistical Review / Revue Internationale de Statistique},
 number = {1},
 pages = {31--36},
 publisher = {[Wiley, International Statistical Institute (ISI)]},
 title = {Orthogonal Polynomial Regression},
 urldate = {2024-11-27},
 volume = {47},
 year = {1979}
}

@article{Yu2005,
  title = {Predictors of response to cardiac resynchronization therapy (PROSPECT)—study design},
journal = {American Heart Journal},
volume = {149},
number = {4},
pages = {600-605},
year = {2005},
issn = {0002-8703},
doi = {https://doi.org/10.1016/j.ahj.2004.12.013},
author = {Cheuk-Man Yu and William T. Abraham and Jeroen Bax and Eugene Chung and Michelle Fedewa and Stefano Ghio and Christophe Leclercq and Angel R. León and John Merlino and Petros Nihoyannopoulos and Dean Notabartolo and Jing {Ping Sun} and Luigi Tavazzi}
}

@article{Zinszer2015,
  title={Forecasting malaria in a highly endemic country using environmental and clinical predictors},
  author={Zinszer, Kate and Kigozi, Ruth and Charland, Katia and Dorsey, Grant and Brewer, Timothy F and Brownstein, John S and Kamya, Moses R and Buckeridge, David L},
  journal={Malaria journal},
  volume={14},
  number={1},
  pages={245},
  year={2015},
  publisher={Springer},
doi={10.1186/s12936-015-0758-4}
}

@book{MacDonald1957,
  title={The epidemiology and control of malaria},
  author={MacDonald, G.},
  year={1957},
  publisher={Oxford University Press},
  address={London}
}

@article{Zinszer2012,
  author = {Zinszer, Kate and Verma, Aman D and Charland, Katia and Brewer, Timothy F and Brownstein, John S and Sun, Zhuoyu and Buckeridge, David L},
	title = {A scoping review of malaria forecasting: past work and future directions},
	volume = {2},
	number = {6},
	year = {2012},
	doi = {10.1136/bmjopen-2012-001992},
	publisher = {British Medical Journal Publishing Group},
journal = {BMJ Open}
}

@book{Anderson1995,
  author    = {James A. Anderson},
  title     = {An Introduction to Neural Networks},
  year      = {1995},
  publisher = {MIT Press},
  address   = {Cambridge, MA}
}

@article{Modeling2019,
 author = {S. Mukhopadhyay and R. Tiwari and P. Shetty and N.J. Gogtay and U.M. Thatte},
title = {Modeling and Forecasting Indian Malaria Incidence Using Generalized Time Series Models},
journal = {Communications in Statistics: Case Studies, Data Analysis and Applications},
volume = {5},
number = {2},
pages = {111--120},
year = {2019},
publisher = {Taylor \& Francis},
doi = {10.1080/23737484.2019.1580629}
}

@techreport{WHO2022,
  title={World Malaria Report},
  author={{World Health Organization}},
  year={2022},
  institution={World Health Organization},
  address={Geneva},
  type={Report}
}

@techreport{WHO2015,
  title={Global technical strategy for malaria 2016--2030},
  author={{World Health Organization}},
  year={2015},
  institution={World Health Organization},
  address={Geneva},
type={Report}
}

@techreport{IndiaFramework2016,
    title={National Framework for Malaria Elimination in India (2016--2030).},
    author={NVBDCP},
    year={2016},
    institution={NVBDCP Delhi},
type={Report}
}

@techreport{IndiaPlan2017,
  title={National Strategic Plan for Malaria 2017--2022},
  author={NVBDCP},
  year={2017},
institution={NVBDCP Delhi},
  type={Report}
}

@book{Hyndman2018,
  title={Forecasting: Principles and Practice},
  author={Hyndman, RJ and Athanasopoulos, G},
  year={2018},
  edition={2},
  publisher={OTexts}
}

@book{box1976time,
  title={Time Series Analysis: Forecasting and Control},
  author={Box, G.E.P. and Jenkins, G.M. and Reinsel, G.C. and Ljung, G.M.},
  year={2015},
  publisher={John Wiley \& Sons}
}

@book{logcorrection,
title={Optimal sports math, statistics, and fantasy},
  author={Kissell, Robert and Poserina, James},
  year={2017},
  publisher={Academic Press}
}

@book{wood2017gam,
  author    = {Wood, Simon N.},
  title     = {Generalized Additive Models: An Introduction with R},
  edition   = {2},
  year      = {2017},
  publisher = {Chapman and Hall/CRC},
  doi       = {10.1201/9781315370279}
}

@article{cleveland1990stl,
  title={S{T}{L}: A seasonal-trend decomposition procedure based on loess},
  author={Robert, Cleveland and William, C and Irma, Terpenning},
  journal={J. Off. Stat},
  volume={6},
  number={1},
  pages={3--73},
  year={1990}
}

@book{draper1998applied,
  title={Applied Regression Analysis},
  author={Norman R. Draper and H. Smith},
  year={1998},
  edition={3},
  publisher={Wiley},
  address={New York}
}

@article{cleveland1979robust,
  author = {William S. Cleveland},
title = {Robust Locally Weighted Regression and Smoothing Scatterplots},
journal = {Journal of the American Statistical Association},
volume = {74},
number = {368},
pages = {829--836},
year = {1979},
publisher = {ASA Website}
}

@book{armstrong1985long,
  author    = {J. Scott Armstrong},
  title     = {Long-Range Forecasting},
  edition   = {2nd},
  year      = {1985},
  publisher = {Wiley},
  address   = {New York}
}

@article{Singh2024,
  title={Time series analysis of malaria cases to assess the impact of various interventions over the last three decades and forecasting malaria in India towards the 2030 elimination goals},
  author={Singh, M.P. and Rajvanshi, H. and Bharti, P.K. and others},
  journal={Malar J},
  volume={23},
  pages={50},
  year={2024},
  doi={10.1186/s12936-024-04872-8}
}

@incollection{Chougale2025,
  title={Predicting Malaria Cases in Mumbai: Insights from Statistical and Machine Learning Models},
  author={Chougale, P.D. and Somaraj, A.B. and Ananthakumar, U.},
  booktitle={Pattern Recognition. ICPR 2024 International Workshops and Challenges},
  editor={Palaiahnakote, S. and Schuckers, S. and Ogier, J.M. and Bhattacharya, P. and Pal, U. and Bhattacharya, S.},
  series={Lecture Notes in Computer Science},
  volume={15615},
  year={2025},
  publisher={Springer, Cham},
  doi={10.1007/978-3-031-87660-8_2}
}

@article{Yadav2022,
  title={An investigation of the efficacy of different statistical models in malaria forecasting in the semi-arid regions of Gujarat, India},
  author={Yadav, Chander Prakash and Baharia, Rajendra and Ranjha, Ritesh and Hussain, Syed Shah Areeb and Singh, Kuldeep and Faizi, Nafis and Sharma, Amit},
  journal={Journal of Vector Borne Diseases},
  volume={59},
  number={4},
  pages={337--347},
  year={2022},
  doi={10.4103/0972-9062.355959}
}

@article{Ghosh2019,
  title={Malaria elimination in India-The way forward},
  author={Ghosh, S.K. and Rahi, M.},
  journal={J Vector Borne Dis},
  volume={56},
  number={1},
  pages={32--40},
  year={2019},
  doi={10.4103/0972-9062.257771},
  pmid={31070163}
}

@article{Dhamnetiya2015,
  title={Malaria Control in India: Strategies, Progress and Challenges},
  author={Dhamnetiya, Deepak and Sahu, Monalisha},
  journal={International Journal of Contemporary Medicine},
  volume={3},
  pages={14},
  year={2015},
  doi={10.5958/2321-1032.2015.00005.4}
}

@techreport{WHO2023,
  title={World Malaria Report 2023},
  author={{World Health Organization}},
  year={2023},
  institution={World Health Organization},
  address={Geneva},
  type={Report}
}

\end{document}